\documentstyle{llncs}

\newtheorem{Definition}{Definition}[section]
\newtheorem{Theorem}{Theorem}[section]
\newtheorem{Lemma}{Lemma}[section]
\newtheorem{Corollary}{Corollary}[section]

\begin{document}

\thispagestyle{empty}

\title{\bf
Quantum finite multitape automata\thanks{e-mail:
ambainis@cs.berkeley.edu, richard.bonner@mdh.se, rusins@cclu.lv,
marats@cclu.lv, marek@cs.bonn.edu}}

\author{
Andris Ambainis,\inst{1}
Richard Bonner,\inst{2}
R\=usi\c n\v s Freivalds,\inst{3}
Marats Golovkins,\inst{3}
and Marek Karpinski\inst{4}}

\institute{
Computer Science Division,
       University of California, Berkeley, CA 94720-2320\thanks{ Supported
by Berkeley Fellowship for Graduate Studies.}\\
\and
Department of Mathematics and Physics, M\" alardalens
University\\
\and
Institute of Mathematics and Computer Science,
 University of Latvia, Rai\c na bulv. 29, Riga, Latvia\thanks{
 Research supported by Grant No.96.0282 from the
 Latvian Council of Science}
\and
Department of Computer Science, University of Bonn,
 53117, Bonn, Germany\thanks{
 Research partially supported by the International Computer Science
 Institute, Berkeley, California, by the DFG grant KA 673/4-1, and
 by the ESPRIT BR Grants 7079 and ECUS030}
}

\date{}

\maketitle

\begin{abstract}
Quantum finite automata were introduced by C. Moore, J. P.
Crutchfield \cite{MC 97}, and by A. Kondacs and J. Watrous \cite{KW
97}. This notion is not a generalization of the deterministic finite
automata. Moreover, in \cite{KW 97} it was proved that not all regular
languages can be recognized by quantum finite automata. A. Ambainis and
R. Freivalds \cite{AF 98} proved that for some languages quantum finite
automata may be exponentially more concise rather than both deterministic
and probabilistic finite automata. In this paper we introduce the notion
of quantum finite multitape automata and prove that there is a language
recognized by a quantum finite automaton but not by deterministic or
probabilistic finite automata. This is the first result on a problem which
can be solved by a quantum computer but not by a deterministic or
probabilistic computer. Additionally we discover unexpected
probabilistic automata  recognizing complicated languages. 
\end{abstract}

\section{Introduction}

Recently a new type of algorithms has appeared, namely, {\em quantum}
algorithms. Nobel prize winner physicist Richard Feynman asked in
\cite{Fe 82} what effects can have the principles of quantum mechanics,
especially, the {\em principle of superposition} on
computation. He gave arguments showing that it might be computationally
expensive to simulate quantum mechanics on classical computers. This
observation immediately lead to a conjecture predicting enormous
advantages to quantum computers versus classical ones. 
D. Deutsch \cite{De 89} introduced the commonly used notion of the
quantum Turing machine and proved that quantum Turing machines compute
exactly the same recursive functions as ordinary deterministic Turing
machines do. When Peter Shor \cite{Sh 94} proved that quantum algorithms
can factorize large integers and compute discrete logarithms in a
polynomial time practical construction of quantum computers became a
problem that involves many people and huge funding. Indeed, building a
quantum computer would be equivalent to building a universal code-breaking
machine since the intractability of the above-mentioned problems is the
fundamental of the public-key cryptography.

Quantum mechanics differs from the classical physics very much. It  
suffices to mention {\em Heisenberg's uncertainty principle} asserting
that one cannot measure both the position and the impulse of a particle
simultaneously precisely. There is a certain trade-off between the
accuracy of the two measurements. Another well-known distinction of
quantum
mechanics from the classical physics is the impossibility to measure any
object without changing the object.

The fundamental atom of information is the quantum bit, henceforth
abbreviated by the term `qbit'.

Classical information theory is based on the
classical bit as
fundamental atom. This classical bit, henceforth called
{\em cbit,} is in one of two
classical states $t$ (often interpreted as ``true'') and $f$ (often
interpreted as ``false'').
In quantum information theory
the most elementary unit of information is
 the {\em quantum bit,}
henceforth called {\em qbit}.
To explain it, we first discuss a {\em probabilistic} counterpart of the 
classical bit, which we call here {\em pbit}. It can be $t$ with a
probability $\alpha $ and $f$ with probability $\beta $, where
$\alpha + \beta = 1$. A {\em qbit} is very much like to {\em pbit} with
the following distinction. For a {\em qbit} $\alpha $ and $\beta $ are
not real but complex numbers with the property
$\|\alpha \|^2 + \|\beta \|^2 = 1$.

Every computation done on qbit s is performed by means of unitary
operators. One of the simplest properties of these operators
shows that such a computation is reversible. The result always determines
the input uniquely. It may seem to be a very strong limitation for such
computations. Luckily this is not so.
It is possible to embed any irreversible computation in an appropriate
environment which makes it reversible. For instance, the computing agent
could keep the inputs of previous calculations in successive order.

The following features of quantum computers are important (but far from
the only characteristic features of them).

\begin{itemize}
\item Input, output, program and memory are represented by qbits.

\item Any computation (step) can be represented by a unitary
transformation of the computer as a whole.

\item Any computation is reversible. Because of the unitarity of the
quantum evolution operator, a deterministic computation can be
performed by a quantum computer if and only if it is reversible.

\item No qbit can be copied. After the qbit is processed, the original
form of it is no more available.

\item Measurements may be carried out on any qbit at any stage of the
computation. However any measurement destroys the information. More
precisely, the measurement turns a qbit into a classical bit with
probabilities dependent on the qbit.

\item Quantum parallelism: during a computation, a quantum computer
proceeds down all coherent paths at once. 

\end{itemize}

Quantum finite automata were introduced twice. First this was done by C.
Moore and J.P.Crutchfield \cite{MC 97}. Later in a different and
non-equivalent way these automata were introduced by A. Kondacs and J.
Watrous \cite{KW 97}.

The first definition just mimics the definition of
1-way finite probabilistic only substituting {\em stochastic} matrices by
{\em unitary} ones. We use a more elaborated definition  \cite{KW 97}.

QFA is a tuple
$M=(Q;\Sigma ;\delta ;q_{0};Q_{acc};Q_{rej})$ where $Q$ is a finite set  
of states, $\Sigma $ is an input alphabet, $\delta$ is a transition
function,
$q_{0}\in Q$ is a starting state, and $Q_{acc}\subset Q$
and $Q_{rej}\subset Q$ are sets of accepting and rejecting states.
The states in $Q_{acc}$ and $Q_{rej}$ are called {\em halting states} and
the states in $Q_{non}=Q-(Q_{acc}\cup Q_{rej})$ are called
{\em non halting states}.
$\kappa$ and $\$$ are symbols that do not belong to $\Sigma$.
We use $\kappa$ and $\$$ as the left and the right endmarker,
respectively. The {\em working alphabet} of
$M$ is $\Gamma = \Sigma \cup \{\kappa ;\$\}$.

A superposition of $M$ is any element of $l_{2}(Q)$ (the space of
mappings from $Q$ to $C$ with $l_{2}$ norm). For $q \in Q$, 
$|q\rangle$ denotes the unit vector which takes value 1 at $q$
and 0 elsewhere. All elements of $l_{2}(Q)$ can be expressed as
linear combinations of vectors $|q\rangle$. We will use $\psi$ to
denote elements of $l_{2}(Q)$.

The transition function $\delta$ maps $Q\times \Gamma \times Q$
to $C$. The value $\delta (q_{1};a;q_{2})$ is the amplitude of  
$|q_{2}\rangle$ in the superposition of states to which $M$ goes
from $|q_{1}\rangle$ after reading $a$. For $a\in \Gamma$, $V_{a}$
is a linear transformation on $l_{2}(Q)$ defined by
$$V_{a}(|q_{1}\rangle)=\sum\limits_{q_{2}\in
Q}\delta(q_{1};a;q_{2})|q_{2}\rangle.$$
We require all $V_{a}$ to be unitary.

The computation of a QFA starts in the superposition $|q_{0}\rangle$.
Then transformations corresponding to the left endmarker $\kappa$,
the letters of the input word $x$ and the right endmarker $\$$ are
applied. The transformation corresponding to $a\in \Gamma$ consists
of two steps.

1. First, $V_{a}$ is applied. The new superposition $\psi^{\prime}$
is $V_{a}(\psi)$ where $\psi$ is the superposition before this step.

2. Then, $\psi^{\prime}$ is observed with respect to the observable
$E_{acc}\oplus E_{rej}\oplus E_{non}$ where
$E_{acc}=span\{|q\rangle :q\in Q_{acc}\}$,   
$E_{rej}=span\{|q\rangle :q\in Q_{rej}\}$,
$E_{non}=span\{|q\rangle :q\in Q_{non}\}$.
This observation gives $x\in E_{i}$ with the probability equal to the
amplitude of the projection of $\psi^{\prime}$.
After that, the superposition collapses to this projection.\\  

If we get $\psi^{\prime} \in E_{acc}$, the input is accepted.
If we get $\psi^{\prime} \in E_{rej}$, the input is rejected.
If we get $\psi^{\prime} \in E_{non}$, the next transformation is
applied.\\

We regard these two transformations as reading a letter $a$.

For probabilistic computation, the property that the probability
of correct answer can be increased arbitrarily is considered evident.
Hence, it is not surprising that \cite{KW 97} wrote
"with error probability bounded away from $1/2$", thinking
that all such probabilities are equivalent.
However, mixing reversible (quantum computation) and non-reversible
(measurements after each step) components in one model
makes it impossible.
This problem was first considered in the paper \cite{AF 98} by A. Ambainis
and R. Freivalds. The following theorems were proved there:

Let $p$ be a prime. We consider the language
$L_p=\{ a^i |\mbox{$i$ is divisible by $p$}\}$.
It is easy to see that any deterministic 1-way finite
automaton recognizing $L_p$ has at least $p$ states.

\begin{Theorem}
\label{T6}
For any $\epsilon>0$, there is a QFA with $O(\log p)$ states recognizing   
$L_p$ with probability $1-\epsilon$.
\end{Theorem}

\begin{Theorem}
\label{T8}
Any 1-way probabilistic finite automaton recognizing $L_p$
with probability $1/2+\epsilon$, for a fixed $\epsilon>0$, has
at least $p$ states.
\end{Theorem}

\begin{Theorem}
\label{C1}
There is a language that can be recognized by a 1-QFA with probability
$0.68...$ but not with probability $7/9+\epsilon$.
\end{Theorem}

We consider only multitape finite automata in this paper. A quantum
automaton is defined in the natural way, demanding that the transformation
( the state, the information on the first head having or not
having moved, the information on the second head having or not having
moved, ... , $\to $ the state, the information on the first head having or
not
having moved, the information on the second head having or not having
moved, ... ) is unitary for arbitrary tuple (the symbol observed by the
first head, the symbol observed by the first head,...).
\begin{Definition}
A quantum finite multitape automaton (QFMA)\\
$A=(Q;\Sigma;\delta;q_0;Q_{acc};Q_{rej})$ is specified by the finite input
alphabet $\Sigma$, the finite set of states $Q$, the initial
state $q_0\in Q$, the sets $Q_{acc}\subset Q$, $Q_{rej}\subset Q$ of
accepting and rejecting states, respectively, with
$Q_{acc}\cap Q_{rej}=\emptyset$, and the transition function
$$
\delta :Q\times \Gamma^m\times\{\downarrow, \to\}^m \longrightarrow C_{[0,1]},
$$
where $m$ is the number of input tapes, $\Gamma =\Sigma\cup\{\kappa,\$\}$
is the tape alphabet of $A$ and $\kappa$,$\$$ are endmarkers not in
$\Sigma$,
which satisfies the following conditions (of well-formedness):\\
1. Local probability condition
$$
\forall(q_1,\sigma)\in Q\times \Gamma^m
\sum\limits_{(q,d)\in Q\times \{\downarrow, \to\}^m}
|\delta(q_1,\sigma,q,d)|=1.
$$
2. Orthogonality of column vectors condition.
$$
\forall q_1,q_2\in Q, q_1\neq q_2, \forall\sigma\in\Gamma^m
\ \ \ \
\sum\limits_{(q,d)\in Q\times \{\downarrow, \to\}^m}
\delta^*(q_1,\sigma,q,d)\delta(q_2,\sigma,q,d)=0.
$$
3. Separability condition.\\
$M=_{def}\{1,2,\ldots,m\}.$
The $k$-th component of an arbitrary vector $s$ will be defined as $s^k$.
We shall understand by $I$ an arbitrary element from the set
$P(M)
-\{\emptyset\}$.
$$
R_I=_{def}A_1\times A_2\times\ldots\times A_m, where
A_i=
\left\{
\begin{array}{l}
\{\downarrow, \to\},\ if\ i\notin I\\
\{"nothing"\},\ if\ i\in I.
\end{array}
\right.
$$
$$
T_I=_{def}B_1\times B_2\times\ldots\times B_m,\ where
\ B_i=
\left\{
\begin{array}{l}
\{\downarrow, \to\},\ if\ i\in I\\
\{"nothing"\},\ if\ i\notin I.
\end{array}
\right.
$$
The function
$R_i\times T_i\stackrel{d_I}{\longrightarrow} \{\downarrow, \to\}^m$
is defined as follows:
$$
d_I^i(r,t)=_{def}
\left\{
\begin{array}{l}
r^i,\ if\ i\notin I\\
t^i,\ if\ i\in I.
\end{array}
\right.
$$

$$
d_I(r,t)=_{def}(d_I^1(r,t),d_I^2(r,t),\ldots,d_I^m(r,t)).
$$

$\forall I\in P(M)-\{\emptyset \}\ \forall \sigma_1\sigma_2\in \Gamma^m
\ \forall q_1,q_2\in Q
\ \forall t_1,t_2\in T_I;$\\
if\ $\forall i\notin I\ \sigma_1^i=\sigma_2^i,\ \forall j \in I
\ t_1^j\neq t_2^j
$
then
$$
\sum\limits_{(q,r)\in Q\times R_I}
\delta^*(q_1,\sigma_1,q,d_I(r,t_1))\delta(q_2,\sigma_2,q,d_I(r,t_2))=0.
$$
\end{Definition}

States from $Q_{acc}\cup Q_{rej}$ are called halting states and states from
$Q_{non}=Q-(Q_{acc}\cup Q_{rej})$ are called non halting states.

To process an input word vector $x\in (\Sigma^*)^m$ by $A$ it is assumed that
the input is written on every tape $k$ with the endmarkers in the form
$w^k_x=\kappa x^k\$$ and that every such a tape, of length $|x^k|+2$,
is circular, i. e., the symbol to the right of $\$$ is $\kappa$.

For the fixed input word vector $x$ we can define $n$ to be an integer
vector which determines the length of input word on every tape.
So for every $n$ we can define $C_n$ to be the set of all possible
configurations of $A$ where $|x^i|=n^i$.
$|C_n|=|Q|\prod\limits_{i=1}^m(n^i+2)$. Every such a configuration
is uniquely determined by a pair $|q,s\rangle$, where
$q\in Q$ and $0\leq s^i\leq |x^i|+1$ specifies the position of head
on the $i$-th tape.

Every computation of $A$ on an input $x$, $|x^i|=n^i$, is specified
by a unitary evolution in the Hilbert space $H_{A,n}=l_2(C_n)$.
Each configuration $c\in C_n$ corresponds to the basis vector in $H_{A,n}$.
Therefore a global state of $A$ in the space $H_{A,n}$ has a form
$\sum\limits_{c\in C_n}\alpha_c|c\rangle$, where
$\sum\limits_{c\in C_n}|\alpha_c|^2=1$.
If the input word vector is $x$ and the automaton $A$ is in its global state
$|\psi\rangle=
\sum\limits_{c\in C_n}\alpha_c|c\rangle$,
then its further step is equivalent to the application of a linear operator
$U_x^\delta$ over Hilbert space $l_2(C_n)$.
\begin{Definition}
$$
U_x^\delta |\psi\rangle=\sum\limits_{c\in C_n}\alpha_cU_x^\delta|c\rangle.
$$
If a configuration $c=|q',s\rangle$, then
$U_x^\delta|c\rangle=
\sum\limits_{(q,d)\in Q\times \{\downarrow, \to\}^m}
\delta(q',\sigma(s),q,d)|q,\tau(s,d)\rangle$,
where $\sigma(s)=(\sigma^1(s),\ldots,\sigma^m(s))$,
$\sigma^i(s)$ specifies the $s^i$-th symbol on the $i$-th tape,
and
$$
\tau(s,d)=(\tau^1(s,d),\ldots,\tau^m(s,d)),
\ \tau^i(s,d)=
\left\{
\begin{array}{l}
(s^i+1)\bmod(n^i+2),\ if\ d^i='\to'\\
s^i,\ if\ d^i='\downarrow'.
\end{array}
\right.
$$
\end{Definition}

\begin{Lemma}
The well-formedness conditions are satisfied iff for any input $x$ the mapping
$U^\delta_x$ is unitary. 
\end{Lemma}

\begin{Definition}
A QFMA $A=(Q;\Sigma;\delta;q_0;Q_{acc};Q_{rej})$ is simple if for each $\sigma \in \Gamma^m$
there is a linear unitary operator $V_\sigma$ over the inner-product space
$l_2(Q)$
and a function $D:Q\longrightarrow\{\downarrow,\to\}^m$, such that
$$
\forall q_1\in Q\ \forall\sigma\in\Gamma^m\ \ \delta(q_1,\sigma,q,d)=
\left\{
\begin{array}{l}
\langle q|V_\sigma|q_1\rangle,\ if\ D(q)=d\\
0,\ otherwise.
\end{array}
\right.
$$
\end{Definition}

\begin{Lemma}
If the automaton $A$ is simple, then conditions of well-formedness are satisfied
iff for every $\sigma$ $V_\sigma$ is unitary.
\end{Lemma}

As in the case of
single-tape quantum finite automata it is presumed that all the states are
divided into {\em halting} and {\em nonhalting}, and whenever, the
automaton comes into a halting state, the automaton stops, and accepts or
rejects the input with a probability equal to the square of the modulo of
the amplitude.

\section{Reversible automata}
\label{S23}

A 1-way reversible finite automaton (RFA) is a QFA with
$\delta(q_1, a, q_2)\in\{0, 1\}$ for all $q_1, a, q_2$.
Alternatively, RFA can be defined as a deterministic automaton
where, for any $q_2, a$, there is at most one state
$q_1$ such that reading $a$ in $q_1$ leads to $q_2$.
We use the same definitions of acceptance and rejection.
States are partitioned into accepting, rejecting and non-halting states
and
a word is accepted (rejected) whenever the RFA
enters an accepting (rejecting) state. After that,
the computation is terminated. Similarly to quantum case,
endmarkers are added to the input word.
The starting state is one, accepting (rejecting) states
can be multiple. This makes our model different from
both \cite{An 82} (where only one accepting state was allowed)
and \cite{Pi 92} (where multiple starting states with a non-deterministic 
choice between them at the beginning were allowed).
We define our model so because we want it to be as close to our
model of QFAs as possible.

Generally, it's hard to introduce probabilism into finite automata
without losing reversibility. However, there are some types
of probabilistic choices that are consistent with reversibility.
For example, it was proved by A. Ambainis and R. Freivalds that for the
language $L=\{ a^{2n+3}|n\in\bbbn\}$ not recognizable by a 1-way RFA,
there are 3 1-way RFAs such that
each word in the language is accepted by 2 of them and  
each word not in the language is rejected by 2 out of 3.

\section{Quantum vs. probabilistic automata}
\label{Section2}

\begin{Definition}
We say that a language $L$ is [m,n]-deterministically recognizable if
there are $n$ deterministic automata $A_1$, $A_2$, $A_n$ such that:\\
a) if the input is in the language $L$, then all $n$
automata $A_1$, \dots , $A_n$ accept the input;\\
b) if the input is not in the language $L$, then at most $m$ of the
automata $A_1$, \dots , $A_n$ accept the input.
\end{Definition}

\begin{Definition}
We say that a language $L$ is [m,n]-reversibly recognizable if
there are $n$ deterministic reversible automata $A_1$, $A_2$, $A_n$ such
that:\\
a) if the input is in the language $L$, then all $n$
automata $A_1$, \dots , $A_n$ accept the input;\\
b) if the input is not in the language $L$, then at most $m$ of the
automata $A_1$, \dots , $A_n$ accept the input.
\end{Definition}

\begin{Lemma}
\label{LQ1}
If a language $L$ is [1,n]-deterministically recognizable by 2-tape finite
automata, then $L$ is recognizable by a probabilistic 2-tape finite
automaton with probability $\frac{n}{n+1}$.
\end{Lemma}

{\bf Proof.}
The probabilistic automaton starts by choosing a random integer $1 \le r
\le (n+1)$. After that , if $r \le n$, then the automaton goes on
simulating the deterministic automaton $A_r$, and, if $r = n+1$, then the
automaton rejects the input. The inputs in $L$ are accepted with
probability $\frac{n}{n+1}$, and the inputs not in the language are
rejected with a probability no less than $\frac{n}{n+1}$.
$\Box$

\begin{Lemma}
\label{LQ2}
If a language $L$ is [1,n]-reversibly recognizable by 2-tape finite
automata, then $L$ is recognizable by a quantum 2-tape finite
automaton with probability $\frac{n}{n+1}$.
\end{Lemma}

{\bf Proof.}
In essence the algorithm is the same as in Lemma \ref{LQ1}.
The automaton starts by choosing a random integer $1 \le r
\le (n+1)$. This is done by taking 3 different actions with amplitudes
$\frac{1}{\sqrt{3}}$ (the possibility to make such a choice is asserted in
Lemma \ref{Q13}). After that , if $r \le n$, then the automaton goes on
simulating the deterministic automaton $A_r$, and, if $r = n+1$, then the
automaton rejects the input. Acceptance and rejecting are made by entering
the states where measurement is made immediately. (Hence the probabilities
are totaled, not the amplitudes.)
$\Box$

First, we discuss the following 2-tape language
$$L_1 = \{ (x_1\nabla x_2, y) \| x_1 = x_2 = y
\},$$
where the words $x_1, x_2, y$ are unary.
  
\begin{Lemma}
\label{LD1}
For arbitrary natural $n$, the language $L_1$ is [1,n]-deterministically
recognizable.
\end{Lemma}

{\bf Proof.} See Appendix.

R. Freivalds \cite{Fr 79} proved

\begin{Theorem}
\label{Fre79}
The language $L_1$ can be recognized
with arbitrary probability $1-\epsilon $ by  a probabilistic 2-tape finite
automaton but this language cannot be recognized by a deterministic 2-tape
finite automaton.
\end{Theorem}

{\bf Proof.}
By Lemma \ref{LD1} $L$ is [1,n]-deterministically recognizable for
arbitrary $n$.By Lemma \ref{LQ1}, the language is recognizable with
probability $\frac{n}{n+1}$.
$\Box$

\begin{Theorem}
\label{T1}
The language $L_1$ can be recognized
with arbitrary probability $1-\epsilon $ by  a quantum 2-tape finite
automaton.
\end{Theorem}

{\bf Proof.}
By Lemma \ref{LQ2}.
$\Box$

We wish to prove a quantum counterpart of Theorem \ref{Fre79}. We need
some lemmas to this goal.

In an attempt to construct a 2-tape language recognizable by a quantum
2-tape finite automaton but not by probabilistic 2-tape finite automata we
consider a similar language
$$L_2 = \{ (x_1\nabla x_2\nabla x_3,y) \| \mbox{there are
exactly 2 values
of} 
x_1 , x_2 , x_3 \mbox{such that they equal} y \},$$
where the words $x_1, x_2, x_3, y$ are unary.

\begin{Theorem}
\label{L_2q}
A quantum automaton exists which recognizes the language $L_2$ with a
probability $\frac{3}{5} - \epsilon $ for arbitrary positive $\epsilon $.
\end{Theorem}

{\bf Proof.}
This automaton
with amplitudes:\\
a) $\frac{1}{\sqrt 5} \times 1$\\
b) $\frac{1}{\sqrt 5} \times (cos \frac{2\pi }{3} + i\sin \frac{2\pi
}{3})$\\
c) $\frac{1}{\sqrt 5} \times (cos \frac{4\pi }{3} + i\sin \frac{4\pi
}{3})$\\
d) $\sqrt \frac{2}{5}$\\
takes actions:\\
a) compare $x_1 = x_2 = y$,\\
b) compare $x_2 = x_3 = y$,\\
c) compare $x_1 = x_3 = y$,\\
d) says "accept".\\
If $y$ equals all 3 words $x_1, x_2, x_3$, then 
the input is accepted with probability $\frac{2}{5}$ (since the
amplitudes of the actions a), b), c) total to $0$).
If $y$ equals 2 out of 3 words $x_1, x_2, x_3$, then
the input is accepted with probability $\frac{3}{5}$.
If $y$ equals at most one of the words $x_1, x_2, x_3$, then
the input is accepted with probability $\frac{2}{5}$ (only if the action
d) is taken).
$\Box$

Unfortunately, the following theorem holds.

\begin{Theorem}
\label{L_2pr}
A probabilistic automaton exists which recognizes the language $L_2$ with
a probability $\frac{21}{40}$
\end{Theorem}

{\bf Proof.}
The probabilistic automaton
with probability $\frac{1}{2}$ takes an  action $A$ or $B$:\\
A) Choose a random $j$ and compare $x_j = y$. If {\bf yes}, accept with
probability $\frac{19}{20}$. If {\bf no}, accept with probability
$\frac{1}{20}$.\\
B) Choose a random pair $j,k$ and compare $x_j = x_k = y$. If {\bf yes},
reject. If {\bf no}, accept with probability
$\frac{12}{20}$.

If $y$ equals all 3 words $x_1, x_2, x_3$
and the action $A$ is taken, then
the input is accepted with relative probability $\frac{19}{20}$.
If $y$ equals all 3 words $x_1, x_2, x_3$, then
and the action $A$ is taken, then
the input is accepted with relative probability $0$.
This gives the acceptance probability in the case
if $y$ equals all 3 words $x_1, x_2, x_3$, to be $\frac{19}{40}$
and the probability of the correct result "no" to be $\frac{21}{40}$.

If $y$ equals 2 words out of $x_1, x_2, x_3$      
and the action $A$ is taken, then
the input is accepted with relative probability $\frac{13}{20}$.
If $y$ equals 2 words out of $x_1, x_2, x_3$
and the action $B$ is taken, then
the input is accepted with relative probability $\frac{8}{20}$.
This gives the acceptance probability in the case
if $y$ equals 2 words out of $x_1, x_2, x_3$, to be $\frac{21}{40}$.

If $y$ equals only 1 word out of $x_1, x_2, x_3$
and the action $A$ is taken, then
the input is accepted with relative probability $\frac{7}{20}$.
If $y$ equals only 1 word out of $x_1, x_2, x_3$
and the action $B$ is taken, then
the input is accepted with relative probability $\frac{12}{20}$.
This gives the acceptance probability in the case
if $y$ equals only 1 word out of $x_1, x_2, x_3$, to be $\frac{19}{40}$
and the probability of the correct result "no" to be $\frac{21}{40}$.
 
If $y$ equals no word of $x_1, x_2, x_3$
and the action $A$ is taken, then
the input is accepted with relative probability $\frac{1}{20}$.
If $y$ equals no word of $x_1, x_2, x_3$
and the action $B$ is taken, then
the input is accepted with relative probability $\frac{12}{20}$.
This gives the acceptance probability in the case
if $y$ equals no word  of $x_1, x_2, x_3$, to be $\frac{13}{40}$
and the probability of the correct result "no" to be $\frac{27}{40}$.
$\Box$

Now we consider a modification of the language $L_2$ which might be more
difficult for a probabilistic recognition:
$$
\begin{array}{l}
L_3 = \{ (x_1\nabla x_2\nabla x_3,y_1\nabla y_2) \| \mbox{there is
exactly one value} k\\ 
\qquad{}\mbox{such that there are exactly two values} j
\mbox{such that} x_j = y_k .
\}
\end{array}
$$

\begin{Theorem}
\label{L_3q}
A quantum finite 2-tape automaton exists which recognizes the language
$L_3$ with a
probability $\frac{6}{11} - \epsilon $ for arbitrary positive $\epsilon $.
\end{Theorem}

{\bf Proof} is moved to Appendix. It is provided for the referees only,
and it will not be included in the final text.

However this language also can be recognized by a probabilistic 2-tape
finite automaton.

\begin{Theorem}
\label{L_3pr}
A probabilistic finite 2-tape automaton exists which recognizes the
language $L_3$ with a
probability $\frac{13}{25} - \epsilon $ for arbitrary positive $\epsilon
$.
\end{Theorem}

{\bf Proof.}
The probabilistic automaton 
with probability $\frac{6}{25}$ takes action $A$ or $B$ or $C$ or with
probability
$\frac{7}{25}$ takes action $D$:\\
A) Choose a random $k$ and two values of $j$. Then compare $x_j = y_k $.
If {\bf yes}, accept. If {\bf no}, reject.\\
B) Chose a random $k$ and compare $x_1 = x_2 = x_3 = y_k$. If {\bf yes},
reject. If {\bf no}, accept.
C) Choose two values $j$ and $m$. Then compare $x_j = x_m = y_1 = y_2 $.  
If {\bf yes}, reject. If {\bf no}, accept.\\     
D) Says "reject".\\

Notice that the actions $A, B, C$ are probabilistic, and they can be
performed only with probability $1-\epsilon $ (actions $A$ and $B$ are
described in the proof of Theorem \ref{Fre79} and action $C$ is similar).

The acceptance probabilities equal:\\
\begin{tabular}{|r||r||r||r||r|}
\hline
& A & B & C & total\\
\hline
no $y_k$ equals 2 or 3 $x_j$ & 0 & 1 & 1 & $\frac{12}{25}$\\
\hline
one $y_k$ equals 2  $x_j $ & $\frac{1}{6}$ & 1 & 1 & $\frac{13}{25}$\\
\hline
one $y_k$ equals 3  $x_j $ & $\frac{1}{2}$ & $\frac{1}{2}$ & 1 &
$\frac{12}{25}$\\
\hline
two $y_k$ equal 2  $x_j $ & $\frac{1}{3}$ & 1 & $\frac{2}{3}$ &
$\frac{12}{25}$\\
\hline
all $y_k$ equal all  $x_j $ & 1 & 0 & 0 & $\frac{6}{25}$\\
\hline
\end{tabular}
$\Box$

Finally we consider a modification of the languages above which indeed is
difficult for a probabilistic recognition:
$$L_4 = \{ (x_1\nabla x_2,y) \| \mbox{there is
exactly one value}\ j
\ \mbox{such that} x_j = y .
\}
$$
where the words $x_1, x_2, y$ are binary.
  
\begin{Theorem}
\label{L_3qqq}  
A quantum finite 2-tape automaton exists which recognizes the language
$L_4$ with a
probability $\frac{2}{3} - \epsilon $ for arbitrary positive $\epsilon $.
\end{Theorem}
  
{\bf Idea of the proof.} The computations corresponding to the checks
whether or not $x_1 = y$ and $x_2 = y$, are performed with opposite
amplitudes. If these two computations are successful, the amplitudes
annihilate.

\centerline{\Huge Appendix}
\centerline{\Large (for referees only)}

\section{Unitary matrices}
\label{Section1}

\begin{Lemma}
\label{Q5}
For arbitrary real values $\phi , \psi , \eta $, the matrix

$$
\left (
\begin{array}{rr}
\cos \phi (\cos \eta +i \sin \eta )& \sin \phi (\cos \eta +i \sin
\eta )\\
\sin \phi (\cos \psi +i \sin \psi )& -\cos \phi (\cos \psi +i \sin \psi )
\end{array}
\right )
$$
is unitary.
\end{Lemma}

\begin{Corollary}
\label{Q6}
The matrix
$\left (
\begin{array}{rr}
\frac{1}{\sqrt{2}}&\frac{1}{\sqrt{2}}\\
\frac{1}{\sqrt{2}}&-\frac{1}{\sqrt{2}}
\end{array}
\right )
$
is unitary.
\end{Corollary}

\begin{Corollary}
\label{Q4}
The matrix
$\left (
\begin{array}{rr}
\cos \phi & i \sin \phi\\
i \sin \phi&\cos \phi
\end{array}
\right )
$
is unitary.
\end{Corollary}

\begin{Corollary}
\label{Q41}
The matrix
$\left (
\begin{array}{rr}
\cos \phi &  \sin \phi\\  
\sin \phi&-\cos \phi
\end{array}
\right )
$
is unitary.
\end{Corollary}

This corollary is crucially important for the sequel. We will use it to
prove that quantum automata (in contrast with deterministic or
probabilistic automata) can do the counting modulo arbitrarily large
prime numbers using only two states.

\begin{Lemma}
\label{Q7}
For arbitrary real values $\phi , \psi $, the matrix

$$
\left (
\begin{array}{rrrr} 
\cos \phi \cos \psi& i\sin \phi \cos \psi& i\cos \phi \sin \psi&
  -\sin \phi \sin \psi \\
 i\sin \phi \cos \psi&\cos \phi \cos \psi&-\sin \phi \sin \psi&
i\cos \phi \sin \psi \\
 i\cos \phi \sin \psi&-\sin \phi \sin \psi &\cos \phi \cos \psi&
i\sin \phi \cos \psi \\
-\sin \phi \sin \psi &i\cos \phi \sin \psi& i\sin \phi \cos \psi&
\cos \phi \cos \psi
\end{array}
\right )   
$$
is unitary.
\end{Lemma}

\begin{Corollary}
\label{Q8}
The matrix
$$
\left (
\begin{array}{rrrr}
\frac{1}{2}& \frac{i}{2}& \frac{i}{2}& -\frac{1}{2}\\
 \frac{i}{2}&\frac{1}{2}&-\frac{1}{2}& \frac{i}{2}\\
 \frac{i}{2}&-\frac{1}{2}&\frac{1}{2}& \frac{i}{2}\\
-\frac{1}{2}& \frac{i}{2}& \frac{i}{2}&\frac{1}{2}
\end{array}
\right )
$$
is unitary.
\end{Corollary}

\begin{Definition}
\label{Q9}
We call the matrix  
$$
C = \left (
\begin{array}{rrrr}
c_{11} & c_{12} & \dots & c_{1 \, kn}\\
c_{21} & c_{22} & \dots & c_{2 \, kn}\\
\dots & \dots & \dots & \dots \\
c_{kn \, 1} & c_{kn \, 2} & \dots & c_{kn \, kn}
\end{array}
\right )
$$
a block-product of the matrices
$
A = \left (
\begin{array}{rrrr}
a_{11} & a_{12} & \dots & a_{1 \, k}\\
a_{21} & a_{22} & \dots & a_{2 \, k}\\
\dots & \dots & \dots & \dots \\
a_{k \, 1} & a_{k \, 2} & \dots & a_{k \, k}
\end{array}
\right )
$
and
$
B = \left (
\begin{array}{rrrr}
b_{11} & b_{12} & \dots & b_{1 \, n}\\
b_{21} & b_{22} & \dots & b_{2 \, n}\\
\dots & \dots & \dots & \dots \\
b_{n \, 1} & b_{n \, 2} & \dots & b_{n \, k}
\end{array}
\right )
$
if $C_{(m-1)k+i \, (l-1)k+j} = a_{i \, j} b_{m \, l}.
$
\end{Definition} 

\begin{Lemma}
\label{Q10}
If the matrices $A$ and $B$ are unitary, then their block-product is also
a unitary matrix.  
\end{Lemma}
 
\begin{Lemma}
\label{Q112}
For arbitrary prime $p$, the matrix
{\tiny
$$
\left ( 
\begin{array}{rrrrr}
\frac{1}{\sqrt{p}} (e^0)&\frac{1}{\sqrt{p}} (e^0)&
\frac{1}{\sqrt{p}} (e^0)& \dots &\frac{1}{\sqrt{p}} (e^0)\\
\frac{1}{\sqrt{p}} (e^{\frac{2p \pi }{p}})&
\frac{1}{\sqrt{p}} (e^{\frac{2(p-1) \pi }{p}})&
\frac{1}{\sqrt{p}} (e^{\frac{2(p-2) \pi
}{p}})& \dots &
\frac{1}{\sqrt{p}} (e^{\frac{2 \pi }{p}})\\
\frac{1}{\sqrt{p}} (e^{\frac{4p \pi }{p}})& 
\frac{1}{\sqrt{p}} (e^{\frac{4(p-1) \pi }{p}})&
\frac{1}{\sqrt{p}} (e^{\frac{4(p-2) \pi }{p}})&
\dots &
\frac{1}{\sqrt{p}} (e^{\frac{4 \pi }{p}})\\
\frac{1}{\sqrt{p}} (e^{\frac{6p \pi }{p}})&
\frac{1}{\sqrt{p}} (e^{\frac{6(p-1) \pi }{p}})&
\frac{1}{\sqrt{p}} (e^{\frac{6(p-2) \pi }{p}})& \dots &
\frac{1}{\sqrt{p}} (e^{\frac{6 \pi }{p}})\\
\dots & \dots & \dots & \dots & \dots \\
\frac{1}{\sqrt{p}} (e^{\frac{(p-1)p \pi }{p}})&
\frac{1}{\sqrt{p}} (e^{\frac{(p-1)(p-1) \pi }{p}})&
\frac{1}{\sqrt{p}} (e^{\frac{(p-1)(p-2) \pi }{p}})& \dots &
\frac{1}{\sqrt{p}} (e^{\frac{(p-1) \pi }{p}})
\end{array}  
\right )
$$
}
is unitary.
\end{Lemma}

\begin{Corollary}
\label{Q12}
For arbitrary prime $p$, there is a unitary matrix $C_p$ of size $p \times
p$ such that all the elements $c_{1j}$ of this matrix equal
$\frac{1}{\sqrt{p}}$.
\end{Corollary}
 
\begin{Corollary}
\label{Q13}
For arbitrary natural number $n$, there is a unitary matrix $C_n$ of size
$n \times
n$ such that all the elements $c_{1j}$ of this matrix equal
$\frac{1}{\sqrt{n}}$.
\end{Corollary}

\begin{Corollary}
\label{Q14}  
For arbitrary natural number $n$, there is a unitary matrix $C_n$ of size
$n \times
n$ such that all the elements $c_{i1}$ of this matrix equal
$\frac{1}{\sqrt{n}}$.
\end{Corollary}

These corollaries are used  as a tool to perform an
equiprobable choice among a finite number of possibilities.
 
\section{Proof of Lemma 3.3}

\begin{Lemma}
\label{LD1}
For arbitrary natural $n$, the language $L_1$ is [1,n]-deterministically
recognizable.
\end{Lemma}

The language $L$ can be recognized by the following team of deterministic
1-way 2-tape finite automata $\{ A_1, A_2, \cdots , A_n \} $.

The automaton $A_r$ performs cycles, each one consisting in reading
$n+1$ digits from $x_1$ and $r$ digits from $y$. When the symbol $\nabla $
is met, the automaton memorizes the remainder of $x_1$ modulo $n$ and goes
on (in cycles) reading $n+1$ digits from $x_2$ and $n+1-r$ digits from
$y$. If the input pair of words is in the language, the processing of the
two tapes takes the same time. In this case the automaton accepts the
pair, otherwise the automaton rejects it. This way, the automaton accepts
the pair of words if and only if
there are nonnegative integers $u,v$ such that:\\
$$
(n+1)u \le x_1
$$
$$
(n+1)(u+1) > x_1
$$
$$
(n+1)v \le x_2
$$
$$
(n+1)(v+1) > x_2
$$
$$
x_1 - (n+1)u = x_2 - (n+1)v = y - ru - (n+1-r)v
$$
If $x_1 = x_2$, then the number $- ru - (n+1-r)v$ does not depend on the
choice of $r$. Either all $x_i$ match the $y$, or no one does. If $x_1
\ne x_2$, then the numbers $- ru - (n+1-r)v$ are
all different for different values of $r$. Hence at most one of them can
match $y$.
$\Box$

\section{Proof of Theorem 3.5}

{\bf Theorem \ref{L_3q}}. A quantum finite 2-tape automaton exists which
recognizes the language
$L_3$ with a
probability $\frac{3}{5} - \epsilon $ for arbitrary positive $\epsilon $.

This automaton takes the following actions
with the following amplitudes:\\
a) With amplitude $\frac{1}{\sqrt{11}} \times (cos \frac{0\pi }{6} + i\,sin
\frac{0\pi }{6})$ compares whether $x_1 =  x_2 = y_1 $;\\
b) With amplitude $\frac{1}{\sqrt{11}} \times (cos \frac{4\pi }{6} + i\,sin
\frac{4\pi }{6})$ compares whether $x_2 =  x_3 = y_1 $;\\
c) With amplitude $\frac{1}{\sqrt{11}} \times (cos \frac{8\pi }{6} + i\,sin
\frac{8\pi }{6})$ compares whether $x_1 = x_3 = y_1 $;\\
d) With amplitude $\frac{1}{\sqrt{11}} \times (cos \frac{6\pi }{6} + i\,sin
\frac{6\pi }{6})$ compares whether $x_1 = x_2 = y_2 $;\\
e) With amplitude $\frac{1}{\sqrt{11}} \times (cos \frac{10\pi }{6} + i\,sin
\frac{10\pi }{6})$ compares whether $x_2 = x_3 = y_2 $;\\
f) With amplitude $\frac{1}{\sqrt{11}} \times (cos \frac{2\pi }{6} + i\,sin
\frac{2\pi }{6})$ compares whether $x_1 = x_3 = y_2 $.\\
g) With amplitude $\sqrt{\frac{5}{11}}$ says "accept".\\
These comparisons are probabilistic actions (as in Theorem \ref{Fre79};  
recall that the words $x_j, y_k$ are unary)
but they are simulated by a quantum automaton. This way, every action is
replaced by several actions the number of which depends on $\epsilon $.
For instance, if $\epsilon = \frac{1}{n}$ then the action a) is replaced
by $n$ actions:\\
a1) With amplitude $\frac{1}{\sqrt{11}n} \times (cos \frac{0\pi }{6} +
i\,sin
\frac{0\pi }{6})$ compares whether
there are nonnegative integers $u,v$ such that:\\
$$
(n+1)u \le x_1
$$
$$
(n+1)(u+1) > x_1
$$
$$
(n+1)v \le x_2
$$
$$
(n+1)(v+1) > x_2
$$
$$
x_1 - (n+1)u = x_2 - (n+1)v = y_1 - u -nv
$$
a2) With amplitude $\frac{1}{\sqrt{11}n} \times (cos \frac{0\pi }{6} +
i\,sin
\frac{0\pi }{6})$ compares whether
there are nonnegative integers $u,v$ such that:\\
$$
(n+1)u \le x_1
$$
$$
(n+1)(u+1) > x_1
$$
$$
(n+1)v \le x_2
$$
$$
(n+1)(v+1) > x_2
$$
$$
x_1 - (n+1)u = x_2 - (n+1)v = y_1 - 2u -(n-1)v
$$
a3) With amplitude $\frac{1}{\sqrt{11}n} \times (cos \frac{0\pi }{6} +
i\,sin
\frac{0\pi }{6})$ compares whether
there are nonnegative integers $u,v$ such that:\\
$$
(n+1)u \le x_1
$$
$$
(n+1)(u+1) > x_1
$$
$$
(n+1)v \le x_2
$$
$$
(n+1)(v+1) > x_2
$$
$$
x_1 - (n+1)u = x_2 - (n+1)v = y_1 - 3u -(n-2)v
$$

---------------\\

an) With amplitude $\frac{1}{\sqrt{11}n} \times (cos \frac{0\pi }{6} +
i\,sin \frac{0\pi }{6})$ compares whether
there are nonnegative integers $u,v$ such that:\\
$$
(n+1)u \le x_1
$$
$$
(n+1)(u+1) > x_1
$$
$$
(n+1)v \le x_2
$$
$$
(n+1)(v+1) > x_2
$$
$$
x_1 - (n+1)u = x_2 - (n+1)v = y_1 - nu -v
$$
If $y_1 = y_2$, then the total of amplitudes for the acceptance is $0$
since the amplitude for comparison of $y_1$ with arbitrary pair $x_i,
x_j$ is (minus 1) times the  amplitude for the comparison of $y_2$ with
the same pair $x_i, x_j$.
  
If $y_1 \ne y_2$, and $y_1 = x_1 = x_2$, then $y_2$ cannot equal more than
one of the $x_j$, namely, $x_3$. In this case, all the actions am) [m =
1,2, \dots , n] end in acception  and so do also no more than one of the
actions bm) ,no more than one of the actions cm), no more than one of the
actions dm), no more than one of the actions em), and no more than one of
the actions fm). The total of the amplitudes for the accepting actions
am) is $\frac{n}{\sqrt{11}n} \times (cos \frac{0\pi }{6} +
i\,sin \frac{0\pi }{6})$

$\Box$

\end{document}